\documentclass[prd,aps,groupedaddress,
showpacs,preprintnumbers,
twocolumn,%
nofootinbib
]{revtex4}%
\usepackage{graphicx}
\usepackage{amsmath}
\usepackage{bm}
\usepackage{epsfig}
\usepackage{amssymb}
\begin{document}
\title{Polarization of prompt 
$\mbox{\boldmath$J/\psi$}$ in proton-proton collisions
at RHIC}
\author{Hee Sok Chung}
\affiliation{Department of Physics, Korea University, Seoul 136-701, Korea}
\author{Seyong Kim}
\affiliation{Department of Physics, Sejong University, Seoul 143-747, Korea}
\affiliation{School of Physics, Korea Institute of Advanced Study,
  Seoul 130-722, Korea}
\author{Jungil Lee}
\affiliation{Department of Physics, Korea University, Seoul 136-701, Korea}
\affiliation{Korea Institute of Science and Technology Information,
Daejeon, 305-806, Korea}
\author{Chaehyun Yu}
\affiliation{Department of Physics, Korea University, Seoul 136-701, Korea}

\date{\today}
\preprint{}
\pacs{12.38.-t}
\begin{abstract}

Within the framework of the nonrelativistic QCD (NRQCD) factorization
approach, we compute the
polarization of prompt $J/\psi$ produced at the Brookhaven's Relativistic
Heavy-Ion Collider from proton-proton collisions at the
center-of-momentum energy $\sqrt{s}=200\,$GeV.
The perturbative contributions are computed at leading order in the
strong coupling constant. The prediction reveals that the color-singlet
contribution severely underestimates the PHENIX preliminary data for
the differential cross section integrated over the rapidity range $|y|<0.35$
and its contribution is strongly transversely polarized, 
which disagrees with the PHENIX preliminary data.
After including the color-octet contributions, we find that the NRQCD
predictions for both the cross section and polarization 
over the transverse-momentum range 
$1.5\,\textrm{GeV}<p_T<5\,\textrm{GeV}$ 
($1.5\,\textrm{GeV}<p_T<2\,\textrm{GeV}$)
integrated over the rapidity range
$|y|<0.35$ ($1.2 < |y| < 2.2$)
agree with the data within errors. 
\end{abstract}
\pacs{13.88.+e,13.85.Ni,13.85.-t}

\maketitle
In the mid-1990s, the Collider Detector at Fermilab (CDF)
Collaboration at the Fermilab Tevatron observed remarkable 
surplus of the \textit{prompt}
$J/\psi$ and $\psi(2S)$ production at large transverse momenta
($p_T$)~\cite{Abe:1997jz,Abe:1997yz}. 
Based on the nonrelativistic quantum chromodynamics (NRQCD) factorization
approach~\cite{Bodwin:1994jh},
Braaten and Fleming explained the large cross section by introducing
the color-octet gluon fragmentation mechanism, which is the NRQCD
extension of the color-singlet version proposed in
Ref.~\cite{Braaten:1993rw}.
According to the 
proposal~\cite{Braaten:1994vv}, the prompt spin-triplet $S$-wave charmonia are 
dominantly
produced from the color-octet spin-triplet $S$-wave charm-anticharm pairs
$[c\bar{c}_8({}^3S_1)]$ that are fragmented from the gluons
at large $p_T$.

Using the approximate heavy-quark spin symmetry of the NRQCD Lagrangian,
Cho and Wise predicted that the spin-triplet $S$-wave charmonia $H$ must be
transversely polarized at large $p_T$~\cite{Cho:1994ih}, if
the color-octet gluon fragmentation 
is the dominant source in that region. 
Subsequently, phenomenological polarization predictions for the prompt 
$\psi(2S)$~\cite{Beneke:1995yb,Beneke:1996yw,Leibovich:1996pa}
and 
$J/\psi$~\cite{Braaten:1999qk,Kniehl:2000nn} at the Tevatron
were carried out. The result is that
these states are the more transversely polarized the larger $p_T$ 
is~\cite{Beneke:1995yb,Beneke:1996yw,Leibovich:1996pa,%
Braaten:1999qk,Kniehl:2000nn}
and the rise of the transverse polarization for the $J/\psi$ is
slower than that of the $\psi(2S)$ because of the feed-downs
from the $P$-wave spin-triplet states $\chi_{cJ}$
for $J=0$, $1$, and $2$~\cite{Braaten:1999qk,Kniehl:2000nn}.
Therefore, the observation of transverse $H$ at large $p_T$
is an important independent confirmation of the color-octet gluon
fragmentation mechanism.

However, the run-I CDF measurement with an integrated luminosity
of about 110\,pb$^{-1}$~\cite{Affolder:2000nn} was in dramatic
contradiction to these predictions. 
To make matters even more confusing,
the run-II CDF data of about
800\,pb$^{-1}$~\cite{Abulencia:2007us} neither agree with the 
predictions nor their own run-I measurement.
The prompt $J/\psi$'s from run-II CDF data
are almost 
unpolarized and even show a
moderate increase in the longitudinal fraction from $p_T=5$
to 30$\,$GeV~\cite{Abulencia:2007us}. 

Further theoretical studies include analyses at higher orders 
in the strong coupling $\alpha_s$ for the 
color-singlet~\cite{Campbell:2007ws,Artoisenet:2007xi,Gong:2008sn,Gong:2008hk%
,Artoisenet:2008zz}
and for the color-octet~\cite{Gong:2008ft} channels
and the relativistic corrections~\cite{Fan:2009zq}.
Even after those corrections, the color-octet mechanism still
contributes to the production rate significantly.
Recent color-singlet-model
calculations
of the polarization of \textit{direct} $J/\psi$ at next-to-leading order
(NLO) in $\alpha_s$ predict strong longitudinal
polarization~\cite{Gong:2008sn,Gong:2008hk},
which disagree with the run-II CDF data over the whole range of $p_T$.
These disagreements have cast doubt on our current
understanding of the charmonium production mechanism
and lead one to analyze hadroproduction of charmonia at
various circumstances.

Recently, the PHENIX Collaboration 
at the Brookhaven's Relativistic Heavy-Ion Collider (RHIC)
reported the cross section~\cite{Adare:2006kf} 
and the polarization~\cite{daSilva:2009yy} of the \textit{inclusive} $J/\psi$
production
in $p p$ collisions at the center-of-momentum (CM) energy 
$\sqrt{s}=200\,$GeV.
The data also
contain the $J/\psi$'s from the $B$ decay, which occupy
a tiny fraction ($0.036^{+0.025}_{-0.023}$)
of the total rate~\cite{Oda:2008zz}.
Therefore, the inclusive charmonium cross section should be
essentially the same as the prompt one.
There are several theoretical analyses on the $J/\psi$ production in $p p$
collisions at RHIC.
Several years ago, Cooper, Liu, and Nayak reported the NRQCD prediction for the
production rate for prompt $J/\psi$~\cite{Nayak:2003jp, Cooper:2004qe}, which
agrees with the run-3 PHENIX data~\cite{Adler:2003qs,deCassagnac:2004kb}.
The authors of Ref.~\cite{Haberzettl:2007kj} explained the
$J/\psi$ production rate with two phenomenological parameters
of a new production mechanism so-called 
$s$-\textit{channel cut}~\cite{Lansberg:2005pc}.
However, the $s$-channel-cut prediction for the
polarization disagrees with the PHENIX preliminary data for
the large-rapidity region $1.2<|y|<2.2$ while 
that for $|y|<0.35$ is consistent with the 
measurement~\cite{Lansberg:2008jn,daSilva:2009yy}. 
Furthermore,
Artoisenet and Braaten~\cite{Artoisenet:2009mk} showed that the
$s$-channel cut can be identified with the charm-pair rescattering and
is not a dominant mechanism for charmonium production in high-energy
collisions. 
Very recently, Brodsky and Lansberg carried out calculations for direct
$J/\psi$ in the color-singlet model at NLO in $\alpha_s$~\cite{Brodsky:2009cf}, 
including contributions from $cg$ fusion. They obtained
a rapidity distribution in good agreement with the 
PHENIX data~\cite{Brodsky:2009cf}
under the assumption that about $40\,\%$ of the $J/\psi$
events come from higher resonances.

In this paper,
we present a quantitative analysis of the cross section and
the polarization of prompt $J/\psi$ produced in $pp$ collisions
at $\sqrt{s}=200\,$GeV
using the NRQCD factorization
formalism~\cite{Bodwin:1994jh}.
The perturbative contributions are computed at leading order (LO) in 
$\alpha_s$. 
By including both the color-singlet and color-octet contributions,
we find that the NRQCD predictions for both the cross section
and polarization agree with the data within errors.

The schematic form of the NRQCD factorization formula for the
differential cross section of the polarized
charmonium $H_\lambda$ with momentum $P$ and spin quantum number $\lambda$ 
is given by
\begin{equation}
d \sigma^{H_\lambda(P)} = d \sigma^{c\bar{c}_{n}(P)}
\langle O_n^{H_\lambda(P)} \rangle, \label{Xsection}
\end{equation}
where the summation is assumed over the index $n$ for 
the color and angular momentum states of the $c\bar{c}$ pair.
The short-distance coefficients $d \sigma^{c\bar{c}_{n}(P)}$,
which are insensitive to the long-distance nature of the quarkonium,
are calculable using perturbative QCD.
The nonperturbative nature of the $H$ is factorized into
the NRQCD matrix elements $\langle O_n^{H_\lambda(P)} \rangle$.
The matrix elements are, in general,
tensors depending on $P$ and the polarization tensor of
the $H_\lambda$. After using the symmetries of NRQCD,
one can reduce these polarized matrix elements 
$\langle O_n^{H_\lambda(P)} \rangle$ in terms of the
scalar matrix elements 
$\langle O_n^{H} \rangle$
that are independent of $P$ and $\lambda$.
The numerical importance of the NRQCD matrix elements can be estimated
based on the velocity-scaling rules of NRQCD~\cite{Bodwin:1994jh}.
For the spin-triplet $S$-wave quarkonia $H=J/\psi$ or $\psi(2S)$,
the most important matrix elements
are $\langle O_1^H (^3S_1)\rangle$ for the color-singlet
state, and $\langle O_8^H (^1S_0) \rangle$, $\langle O_8^H (^3S_1)
\rangle$, and $\langle O_8^H (^3P_0) \rangle$ for the color-octet
states, respectively. Because prompt $J/\psi$'s include the samples
that come from the decay of $\psi(2S)$ and 
$\chi_{cJ}$, we also have to consider
the production of $\chi_{cJ}$.  For the $\chi_{cJ}$, 
the
matrix elements
$\langle O_1^{\chi_{c0}} (^3P_0)\rangle$ and 
$\langle O_8^{\chi_{c0}}(^3S_1) \rangle$
are equally important.

In the PHENIX experiment of $p p$ collisions,
they probe a moderate $p_T$ range,
where the fragmentation effect is negligible~\cite{Cooper:2004qe}. Therefore,
the production of $H$ should be dominated by the fusion contributions. 
Then the differential cross section for $H_\lambda$ is expressed as
\begin{equation}
d \sigma^{H_\lambda(P)}
=
f_{i/p}\otimes f_{j/p}\otimes 
d \hat{\sigma}^{c\bar{c}_{n}(P)}_{ij}
\langle O_n^{H_\lambda(P)} \rangle, \label{Xsection2}
\end{equation}
where $f_{i/p}(x,\mu)$ is the parton distribution function (PDF) and
the sums over the partons $i$ and $j$ are implied.
Here, $x$ and $\mu$ are the longitudinal momentum fraction of the
parton and the factorization scale, respectively.

We proceed to compute the differential cross section for the $H_\lambda$.
In order to make a prediction that is consistent with that for
the Tevatron~\cite{Braaten:1999qk}, we follow the strategies
of Ref.~\cite{Braaten:1999qk} except that we neglect
the fragmentation. We include the parton processes
$ij\to c\bar{c}+k$, with $i,\,j=g,\,q,\,\bar{q}$ and $q=u,\,d,\,s$,
and neglect heavy partons like $c$ and $b$.
Corresponding LO parton cross sections $d\hat{\sigma}$ of
order $\alpha_s^3$ are given
in Refs.~\cite{Leibovich:1996pa,Beneke:1998re,Kniehl:2000nn}.
The numerical values for the relevant NRQCD matrix elements
are given in Table I of Ref.~\cite{Braaten:1999qk}.\footnote{
The color-octet matrix elements $\langle O_8^H (^1S_0) \rangle$
and $\langle O_8^H (^3P_J) \rangle$ are determined not separately but 
as a linear combination 
$M_r^H=\langle O^H_8(^1S_0)\rangle + r \langle O^H_8 (^3P_0) \rangle/m_c^2$ 
which depends on the variable $r$ given in Ref.~\cite{Braaten:1999qk}.}
For the PDF's, we choose MRST98LO~\cite{Martin:1998sq} as the default
value and CTEQ5L~\cite{Lai:1999wy} for comparison.
We use the transverse mass $m_T=(4 m_c^2 + p_T^2)^{1/2}$ as
a common scale for the factorization scale and the renormalization scale
with the charm-quark mass $m_c=1.5$ GeV.
We evaluate $\alpha_s$
from the one-loop formula
using the value of $\Lambda_{\rm QCD}$ given in each PDF 
set~\cite{Martin:1998sq,Lai:1999wy}. 
We estimate theoretical uncertainties in our numerical
calculations following Ref.~\cite{Braaten:1999qk}. The errors are from
the matrix elements in Table I of Ref.~\cite{Braaten:1999qk}
and from
the variations of $\mu$ and $m_c$ within the ranges $\frac{1}{2} m_T$--$2 m_T$
and $1.45$--$1.55$ GeV with the central values $m_T$ and $1.5$ GeV,
respectively. The errors in PDF are estimated by taking the
difference between MRST98LO and CTEQ5L. 
We put the color-octet ${}^1S_0$ and ${}^3P_0$ matrix elements as 
$\langle O^H_8(^1S_0) \rangle=xM_r^H$, 
$\langle O^H_8(^3P_0) \rangle/m_c^2=(1-x)M_r^H/r$
and vary $x$ from 0 to 1 with the central value  $\frac{1}{2}$.
All of the errors listed above are added in quadrature.

One of the most convenient measures of the polarized cross section
for the spin-triplet $S$-wave quarkonium $H$ is the variable
$\alpha$ defined by 
\begin{equation}
\alpha = (\sigma_T-2\sigma_L)/(\sigma_T+2\sigma_L),
\label{alpha}%
\end{equation}
where $\sigma_T$ ($\sigma_L$) is the cross section for the 
transversely (longitudinally) polarized $H$.
For the complete transverse (longitudinal) case, 
$\alpha=+1$ ($-1)$. If the $H$ is unpolarized, then
$\alpha=0$.\footnote{The polarized cross section varies depending on the
spin quantization axis~\cite{Braaten:2008xg,Braaten:2008mz}.} 
In the recent PHENIX analysis~\cite{daSilva:2009yy}, 
they used the hadron CM frame like
the CDF analyses~\cite{Abulencia:2007us,Adare:2006kf}.
Employing the method given in Refs.~\cite{Braaten:1999qk,Kniehl:2000nn},
one can compute the polarized cross sections $\sigma_T$ and $\sigma_L$
in the hadron CM frame from Eq.~(\ref{Xsection2}).
Substituting these results to Eq.~(\ref{alpha}), we obtain $\alpha$.

\begin{figure}
\epsfig{file=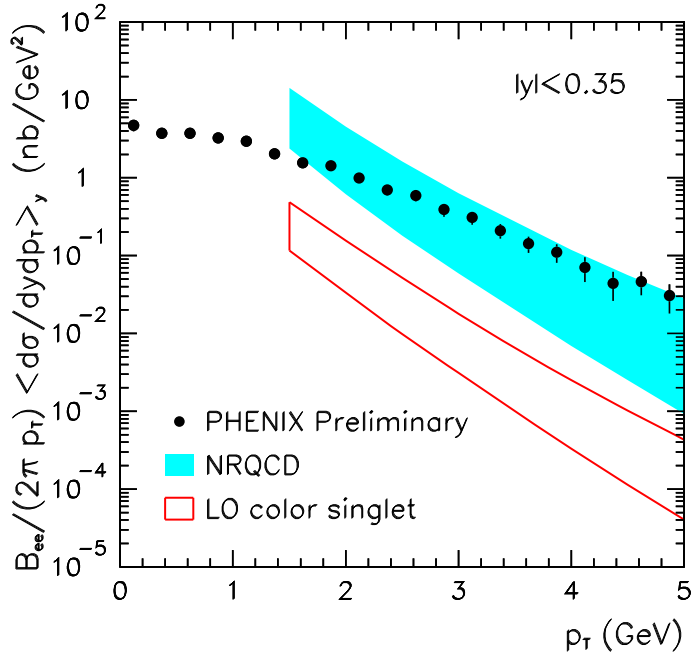,width=10.0cm}
\caption{The differential cross section of prompt $J/\psi$
in $pp$ collisions at $\sqrt{s}=200\,$GeV
as a function of $p_T$ in units of nb/GeV$^2$ against the PHENIX
preliminary
data~\cite{daSilva:2009yy} for the inclusive $J/\psi$.
The shaded band represents the NRQCD prediction and the band
surrounded by a solid curve is the color-singlet contribution
at LO in $\alpha_s$.  
\label{fig:sigma}
}
\end{figure}
Our prediction for the differential cross section for the prompt
$J/\psi$ is shown in Fig.~\ref{fig:sigma} as a function of $p_T$
against the PHENIX preliminary 
data~\cite{daSilva:2009yy}.\footnote{The PHENIX preliminary data quoted in
Fig.~\ref{fig:sigma} do not include an additional global systematic error of
$10\,\%$.}
The rate is averaged over the midrapidity region and
its explicit normalization is given by
\begin{equation}
\frac{B_{ee}}{2\pi p_T}
\left\langle 
{d^2\sigma}/{dy\,dp_T}
\right\rangle_y
=
\frac{B_{ee}}{2\pi p_T(2Y)}\int_{-Y}^{Y}\frac{d^2\sigma}{dydp_T}\,dy,
\label{fofpt}%
\end{equation}
where $Y=0.35$ is the rapidity cut and $B_{ee}$ is the branching
fraction for $J/\psi\to e^+e^-$. The shaded band indicates the NRQCD
prediction at order $\alpha_s^3$. 
The dominant source of the theoretical
uncertainties is from the scale $\mu$, which produces errors of
about $78\,\%$ ($94\,\%$) at $p_T=1.5\,$GeV ($5\,$GeV). 
The variation of $m_c$ produces errors of about $11\,\%$ ($2\,\%$)
at $p_T = 1.5\,$GeV ($5\,$GeV). The
remaining contributions from the matrix elements, $x$, and PDF's
produce errors of about $11\,\%$ ($4\,\%$) at $p_T=1.5\,$GeV
($5\,$GeV).

In the region $p_T>1.5\,$GeV, the NRQCD prediction agrees with the data
within theoretical uncertainties. The curve for the central
value tends to overestimate (underestimate)
the data in the region $p_T < 3\,$GeV ($> 3\,$GeV).
In Fig.~\ref{fig:sigma}, the color-singlet contribution at LO
in $\alpha_s$ is displayed as a band surrounded by a solid curve, 
which severely underestimates the data over the whole range of $p_T$.
In the lower $p_T$ region, the fixed-order calculation of order
$\alpha_s^{3}$ fails to give reliable predictions. In order to extend
the prediction to the lower $p_T$ region, one must include the 
order-$\alpha_s^2$ $2\to 1$ parton processes, its NLO
contribution of order $\alpha_s^3$, and multiple soft-gluon
emissions that should be resummed,\footnote{
See, for example, Ref.~\cite{Bodwin:2005hm,Kniehl:2006sk} 
and references therein.} 
which are out of the scope of this work.

\begin{figure}
\epsfig{file=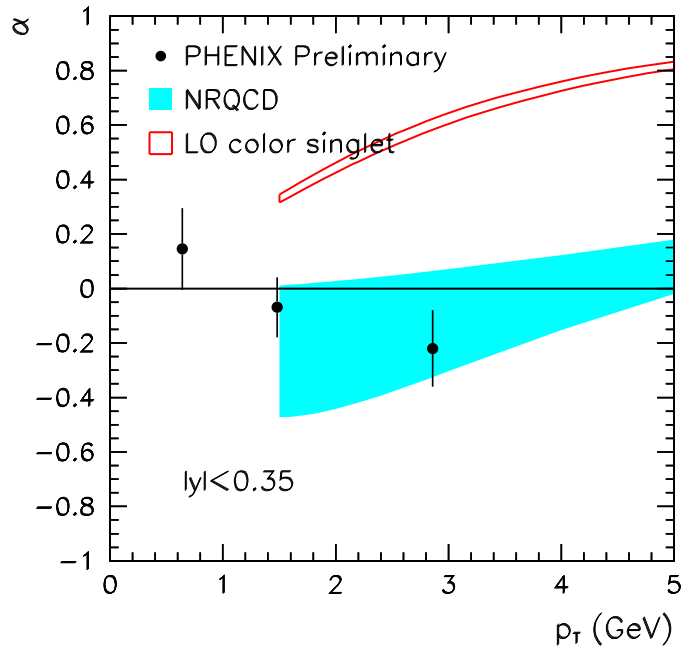,width=10.0cm}
\caption{The polarization parameter $\alpha$ for the
prompt $J/\psi$ in $pp$ collision at $\sqrt{s}=200\,$GeV
as a function of $p_T$ against the PHENIX preliminary
data~\cite{daSilva:2009yy} for the inclusive $J/\psi$.
The shaded band represents the NRQCD prediction and the band
surrounded by a solid curve is the color-singlet contribution
at LO in $\alpha_s$.  \label{fig:alpha}}
\end{figure}

Our results for the $\alpha$ of the
prompt $J/\psi$ integrated over the midrapidity region $|y| < 0.35$ is
shown in Fig.~\ref{fig:alpha} against
the PHENIX preliminary data~\cite{daSilva:2009yy}.
The shaded band represents the NRQCD prediction and the band
surrounded by a solid curve is that for the color-singlet contribution.
The uncertainties of $\alpha$ are computed in the same manner as those of the
differential cross section in Fig.~\ref{fig:sigma}. 
In this case, a large portion of the uncertainties cancel because 
$\alpha$ is computed
as the ratio of polarized cross sections. In contrast to the cross section,
the dominant sources of the errors are the matrix elements, $x$, and PDF's that
account for $99\,\%$ in total.

We predict that the prompt $J/\psi$ is almost unpolarized
or slightly longitudinally polarized in the range 
$1.5\,\textrm{GeV}<p_T<5\,$GeV.
Although the NRQCD prediction has large theoretical uncertainties,
it agrees with the two data points within errors.
As we have expected, 
the fragmentation dominance does not occur and, therefore,
a strong transverse polarization is not observed.
The color-singlet contribution is strongly
transversely polarized with small uncertainties, which is disfavored
by the data.

The polarization variable $\alpha$ for the large-rapidity region $1.2 < |y| <
2.2$ was also measured by the PHENIX Collaboration at $p_T = 1.6\,$GeV as
$\alpha = 0.02 \pm 0.16$~\cite{daSilva:2009yy}. 
Our predictions of the color-singlet model and NRQCD
are $\alpha_{\rm color-singlet} = 0.19 \pm 0.03$ and 
$\alpha_{\rm NRQCD} = 0.15 \pm 0.10$,
respectively. 
On the other hand, one can compare the differential cross section in the 
large-rapidity range near $p_T = 1.6\,$GeV. 
We find that $f(1.6\,{\rm GeV})_{\rm color-singlet} 
= 0.08 \pm 0.05\,{\rm nb/GeV}^2$
and $f(1.6\,{\rm GeV})_{\rm NRQCD} = 2.24 \pm 1.82 \,{\rm nb/GeV}^2$,
where $f(p_T)$ is the same as the right side of Eq.~(\ref{fofpt}) except that
the average is over the range $1.2 < |y| < 2.2$. 
The color-singlet model prediction severely underestimates the differential 
cross
section and the NRQCD prediction agrees with data in both $\alpha$ and $f(p_T)$ in the large-rapidity region.

Our results in Fig.~\ref{fig:alpha} are 
the first NRQCD predictions that can be compared with
the PHENIX results for the $p_T$ distributions of 
the polarization~\cite{daSilva:2009yy}. 
Our $p_T$ distribution for the production rate shown in Fig.~\ref{fig:sigma}
integrated over the midrapidity range agrees with a previous result in
Ref.~\cite{Cooper:2004qe} within errors.
Very recently, Brodsky and Lansberg computed
the rapidity distribution of $J/\psi$ integrated over $p_T$
within the color-singlet model at NLO in $\alpha_s$.
Unfortunately, we are unable to compare our results directly with those in
Ref.~\cite{Brodsky:2009cf} because they did not calculate the $p_T$
distribution and because our LO calculation breaks down at small $p_T$ so we
cannot calculate the rapidity distribution integrated over the whole $p_T$
range.

In conclusion, we have provided the NRQCD predictions for 
the $p_T$ distribution of the differential cross section 
and the polarization variable $\alpha$
for the prompt $J/\psi$ produced in $p p$ collisions at
$\sqrt{s}=200\,$GeV. The short-distance processes were 
computed at the fixed-order $\alpha_s^3$.
We have chosen the numerical values for the
nonperturbative NRQCD matrix elements that were fit
to the prompt $J/\psi$ cross section measured at the Tevatron
and were used to compute the NRQCD prediction for the
$\alpha$ at the Tevatron.
The prediction reveals that the color-singlet
contribution severely underestimates the data for
the differential cross section. 
The NRQCD predictions
for both the cross section and polarization agree with
the data within errors over the range
$1.5\,\textrm{GeV}<p_T<5\,\textrm{GeV}$ 
($1.5\,\textrm{GeV}<p_T<2\,\textrm{GeV}$)
integrated over the rapidity range
$|y|<0.35$ ($1.2 < |y| < 2.2$).
Our fixed-order calculation breaks down for $p_T$ below $1.5\,$GeV.
In order to extend the NRQCD prediction to the lower $p_T$ region, 
one must include the $2 \to 1$ processes at NLO accuracies 
and resum large logarithms from soft-gluon emissions.
It would be exciting to see whether the future NRQCD prediction 
at this region agrees
with the measured PHENIX preliminary data at the lowest $p_T$ bin.

\begin{acknowledgments}

We express our gratitude to Marzia Rosati and Cesar Luiz da Silva 
for drawing our attention to the problem discussed in this paper 
and providing us with useful information regarding the PHENIX experiment.
We are grateful to Eric Braaten for his valuable comments and careful reading 
of the manuscript.
This work was supported by the Korea Research Foundation Grant funded
by the Korean Government (MOEHRD, Basic Research Promotion Fund)
(KRF-2006-C00020). 
H.S.C. and J.L. were supported by the 
Basic Science Research Program of the National Research Foundation of Korea
under Contract No. KRF-2008-313-C00163.
The work of C.Y. was supported 
by Basic Science Research Program through the National Research Foundation of
Korea (NRF) funded by the Ministry of Education, Science and Technology
(2009-0072689).
\end{acknowledgments}


\end{document}